\documentclass{elsart}

\usepackage{epsfig}

\usepackage{amssymb}

\begin{document}

\begin{frontmatter}



\title{Three-Particle Azimuthal Correlations}


\author{Jason Glyndwr Ulery (for the STAR Collaboration)}

\address{Department of Physics, Purdue University, West Lafayette, IN 47907, USA}

\begin{abstract}
% Text of abstract
Two-particle azimuthal correlations reveal broadened and softened away-side correlations. Several different physics mechanisms are possible: large angle gluon radiation, deflected jets, and conical flow or Cerenkov radiation.  Three-particle correlations are investigated to try to discriminate these mechanisms.  We present results on 3-particle azimuthal correlations between a trigger particle of $3<p_T<4$ GeV/c and two softer particles of $1<p_T<2$ GeV/c for $pp$, d+Au and Au+Au collisions at $\sqrt{s_{NN}}=200$ GeV. Implications of the results are discussed.
\end{abstract}

\begin{keyword}
Heavy-ion \sep Azimuthal correlation \sep Three-particle \sep Mach-cone 
\PACS 25.75.-q \sep 25.75.Dw
\end{keyword}
\end{frontmatter}

% main text
\section{INTRODUCTION}
Jets and jet-correlations are good probes to study the medium created in relativistic heavy-ion collisions because their properties in vacuum can be calculated by perturbative quantum chromodynamics.  Two-particle azimuthal correlations with a high-$p_{T}$ trigger particle have shown broadened or even double-humped structures on the away side in central Au+Au collisions \cite{b1} (see Fig.~\ref{fig:Fig2}a).  The shape on the away side is consistent with several physics mechanisms:  large angle gluon radiation \cite{b3}, jets deflected by radial flow or preferential selections of particles due to path-length dependent energy loss, hydrodynamic conical flow generated by Mach-cone shock waves \cite{b4}, or Cerenkov gluon radiation \cite{bx}.  We present a 3-particle correlation analysis designed to differentiate mechanisms with conical emission from the others.

\section{ANALYSIS}

\begin{figure}[htbp]
	\centering
	%\vspace*{-.50cm}
		\includegraphics[width=0.99\textwidth]{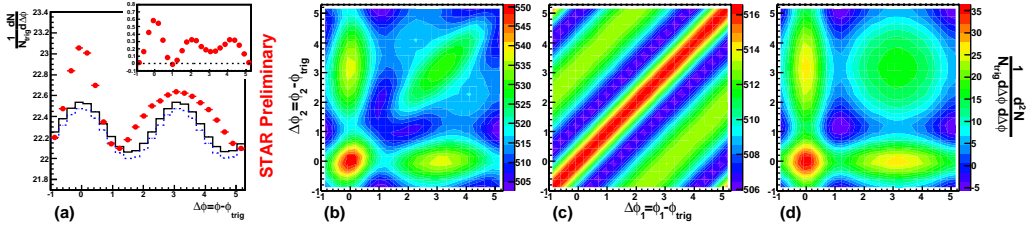}
			\vspace*{-0.1cm}
	\caption{(color online) (a) Raw 2-particle correlation (points), background from mixed events with flow modulation added-in (solid) and scaled by ZYA1 (dashed), and background subtracted 2-particle correlation (insert).  (b) Raw 3-particle correlation, (c) soft-soft background, $ba^{2}B_{3}^{mb}$ and (d) hard-soft background + trigger flow, $\hat{J}_{2}\otimes aB_{2}^{mb}$ + $ba^{2}B_{3}^{mb,TF}$.  See text for detail.  Plots are from ZDC-triggered 0-12\% Au+Au collisions.}
	\label{fig:Fig2}
\end{figure}

The 3-particle jet-correlation analysis method is described in detail in \cite{method}.  The results reported here are between a trigger charged particle with $3<p_T<4$ GeV/c and two associated charged particles of $1<p_T<2$ GeV/c measured by the STAR TPC.  Figure~\ref{fig:Fig2}b displays the raw 3-particle azimuthal correlation.  
Backgrounds must be subtracted to extract the genuine 3-particle jet-correlation signal. 
In one background, called hard-soft background, one associated particle is jet-like correlated with the trigger and the other is not.  It is constructed from the 2-particle jet-like correlation, $\hat{J}_{2}$, folded with the 2-particle background, $B_{2}^{mb}$.  The 2-particle background is constructed by mixed events with the flow modulation added in pairwise from the average $v_2$ of the measured reaction plane \cite{b5} and 4-particle results \cite{b6} and $v_{4}=1.15v_{2}^{2}$ from a fit to data \cite{b5}.  It is normalized (with scale factor $a$) to the signal within $0.8<|\Delta\phi|<1.2$ (zero yield at 1 radian or ZYA1).  We shall refer to the hard-soft background as $\hat{J}_{2}\otimes aB_{2}^{mb}$.

In another background, both associated particles are uncorrelated with the trigger (soft-soft background).  It is obtained by mixing the trigger with inclusive events (minimum bias events from the same centrality), $B_{3}^{mb}$.  This contains all of the correlations between the two associated particles, since they are from the same inclusive event, that are independent of the trigger, including minijets and flow.
Correlations due to anisotropic flow between the trigger and the associated particles, $B_{3}^{mb,TF}$, are added in triplet-wise by mixing the trigger with two different inclusive events.  The total background is $\hat{J}_{2}\otimes aB_{2}^{mb}$ + $ba^{2}(B_{3}^{mb}+B_{3}^{mb,TF})$ where $b$ takes into account the possible differences in the number of pairs in the inclusive event and the underlying event.  The normalization factor $b$ is obtained such that the projection of the 3-particle correlation to either $\Delta\phi$ axis is ZYA1, in the same manner as the 2-particle correlations.  Figure~\ref{fig:Fig2}c shows $ba^{2}B_{3}^{mb}$ and Figure~\ref{fig:Fig2}d shows $\hat{J}_{2}\otimes aB_{2}^{mb}$ + $ba^{2}B_{3}^{mb,TF}$.  

The major sources of systematic uncertainties are the elliptic flow measurements and background normalization.  Other sources include the effect on the trigger particle flow from requiring a correlated particle, uncertainty in the $v_{4}$ parameterization, and multiplicity bias effects on the soft-soft background.
\section{RESULTS}

\begin{figure}[htbp]
	\centering
	%\vspace*{-.50cm}
		\includegraphics[width=0.91\textwidth]{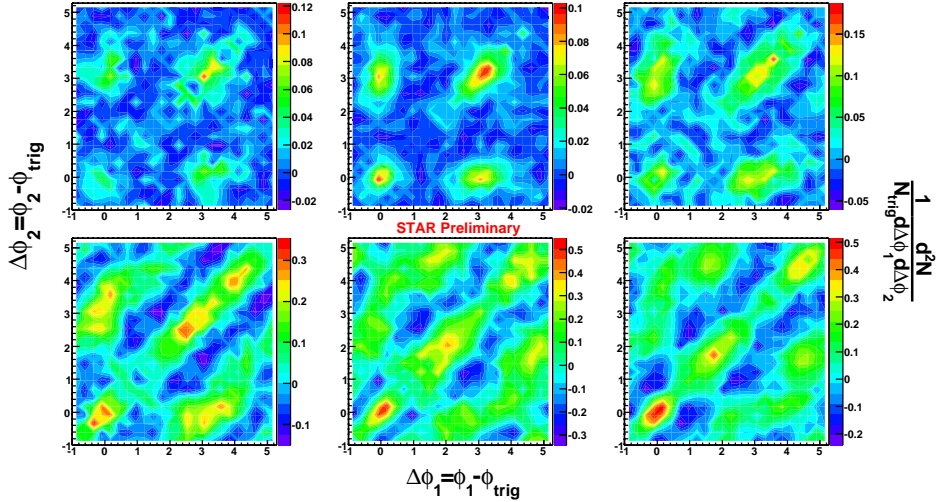}
			\vspace*{-0.5cm}
	\caption{(color online) Background subtracted 3-particle correlations for {\it pp} (top left), d+Au (top middle), and Au+Au 50-80{\%} (top right), 30-50{\%} (bottom left), 10-30{\%} (bottom center), and ZDC triggered 0-12{\%} (bottom right).}
	\label{fig:Fig4}
\end{figure}

The background subtracted 3-particle correlation results are shown in Figure~\ref{fig:Fig4}.  The {\it pp}, d+Au and peripheral 50-80\% Au+Au results are similar.  Peaks are clearly visible for the near-side, the away-side and the two cases of one particle on the near-side and the other on the away-side.  The peak at ($\pi$,$\pi$) displays a diagonal elongation, consistent with $k_T$ broadening.  The additional broadening in Au+Au may be due to deflected jets.  The more central Au+Au collisions display off-diagonal structure, at about $\pi\pm1.3$ radian, that is consistent with conical emission.  The structure increases in magnitude with centrality and is quite clear in the high statistics top 12\% central data afforded by the on-line ZDC trigger.
\label{}

\begin{figure}[htbp]
\hfill
\begin{minipage}[t]{.44\textwidth}
	\centering
	%\vspace*{-.50cm}
		\includegraphics[width=1.0\textwidth]{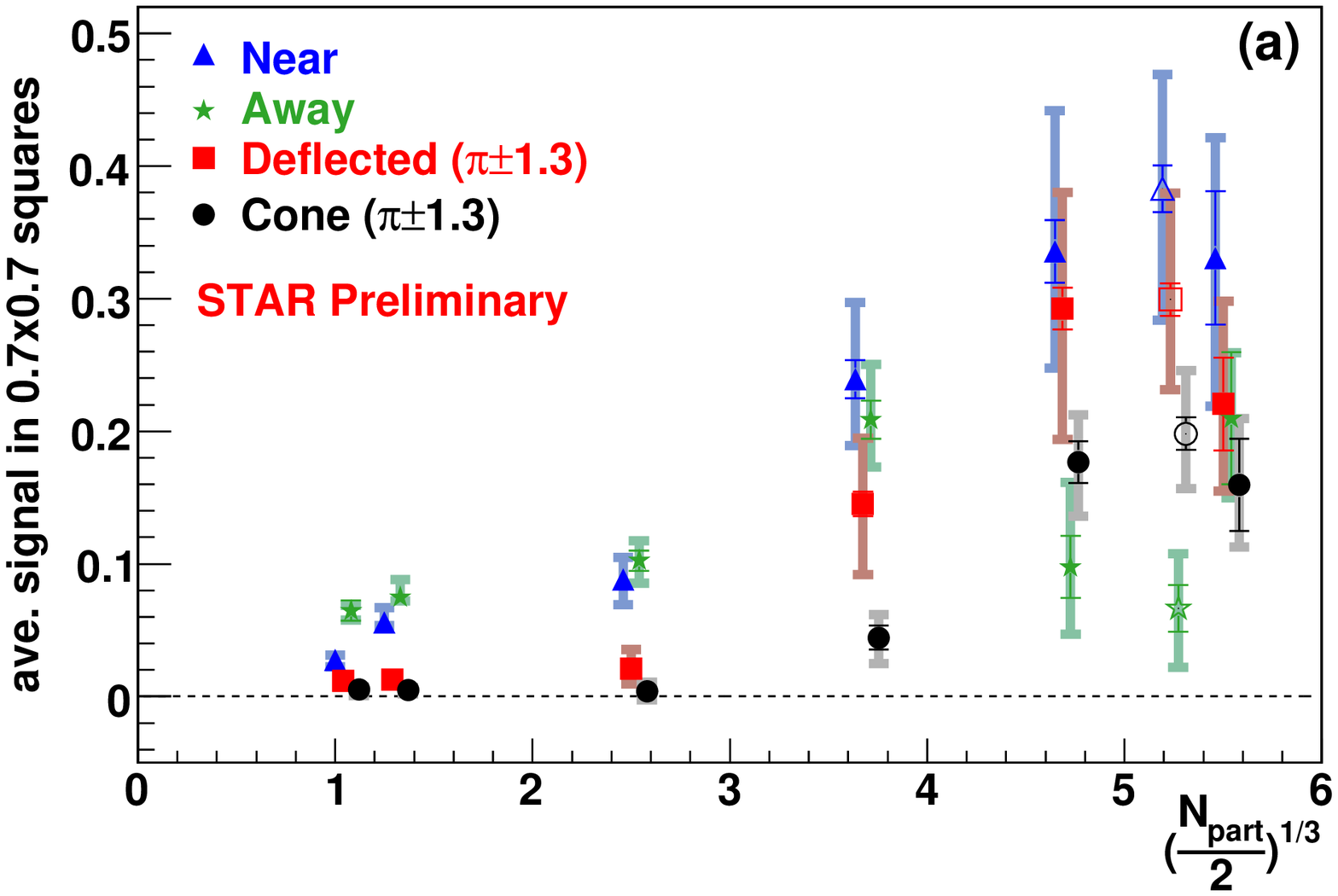}
			\end{minipage}
	\hfill
\begin{minipage}[t]{.44\textwidth}
	\includegraphics[width=1.0\textwidth]{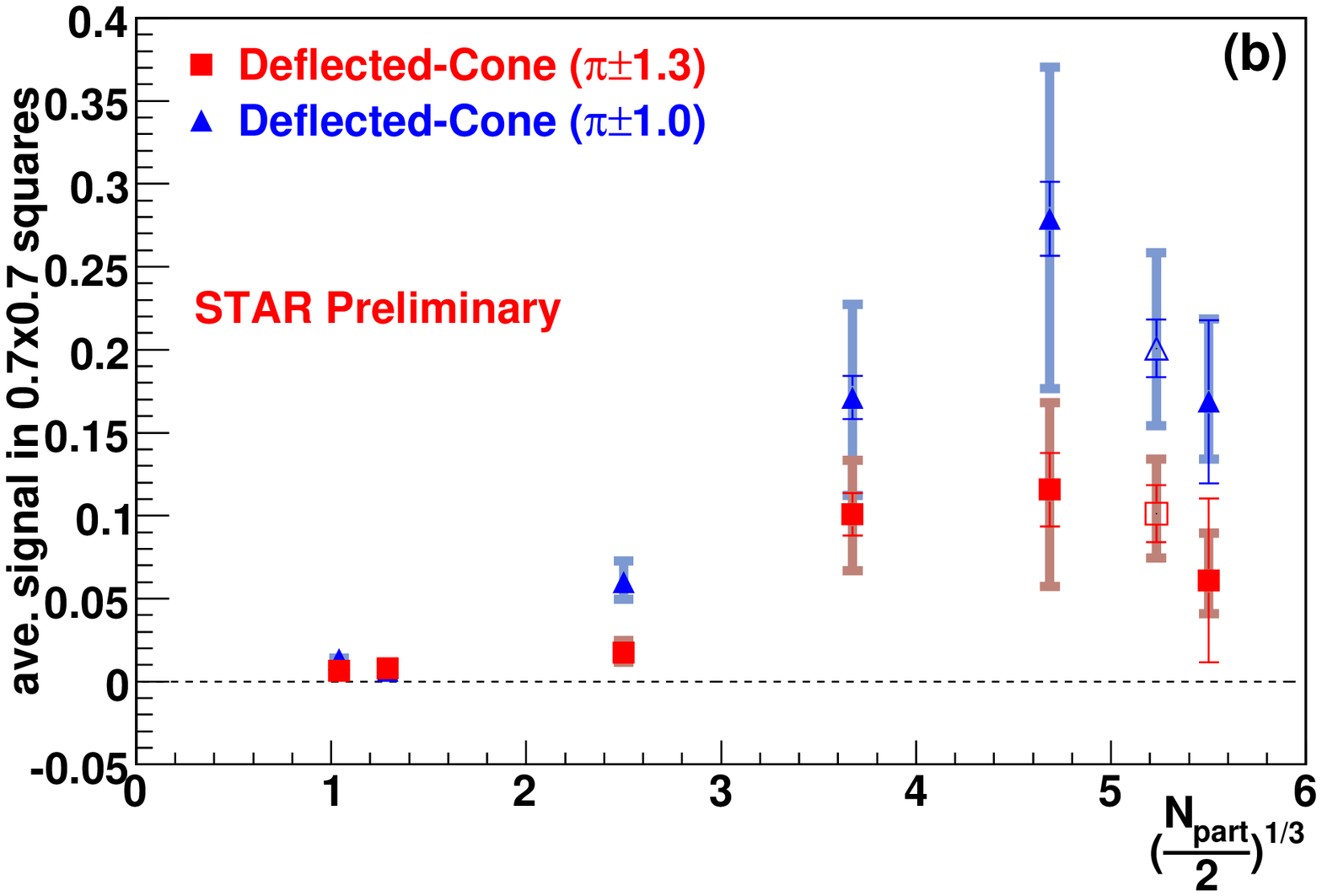}
			\end{minipage}
	\hfill
			\vspace*{-0.4cm}
	\caption{(color online) (a) Average signals in 0.7 $\times$ 0.7 boxes at (0,0) (triangle), ($\pi$,$\pi$) (star), ($\pi\pm1.3$,$\pi\pm1.3$) (square), and ($\pi\pm1.3$,$\pi\mp1.3$) (circle).  (b) Differences between average signals, between ($\pi\pm1.3$,$\pi\pm1.3$) and ($\pi\pm1.3$,$\pi\mp1.3$) (square), and between ($\pi\pm1.0$,$\pi\pm1.0$) and ($\pi\pm1.0$,$\pi\mp1.0$) (triangle).  Solid error bars are statistical and shaded are systematic.  $N_{part}$ is the number of participants.  The ZDC 0-12\% points (open symbols) are shifted to the left for clarity.}
	\label{fig:Fig5}
\end{figure}
\label{}

Figure~\ref{fig:Fig5}a shows the centrality dependence of the average signal strengths in different regions.  The off-diagonal signals (circle) increase with centrality and significantly deviate from zero in central Au+Au collisions.  
Figure~\ref{fig:Fig5}b shows the differences between on-diagonal signals, where both conical emission and deflected jets may contribute, and off-diagonal signals, where only conical emission contributes.  Since conical emission signals are of equal magnitude on-diagonal as off-diagonal, the difference may indicate the contribution from deflected jets.  The difference decreases with distance from ($\pi$,$\pi$).

\section{CONCLUSION}
Three-particle azimuthal correlations have been studied for $3<p_T<4$ GeV/c trigger particles and $1<p_T<2$ GeV/c associated particles.  Au+Au, d+Au and {\it pp} collisions at $\sqrt{s_{NN}}$=200 GeV are compared.  The centrality dependence of Au+Au collisions is analyzed.  Diagonal elongation is seen in {\it pp} and d+Au possibly due to $k_{T}$ broadening.  Further elongation in Au+Au may indicate additional contribution from deflected jets.  Off-diagonal peaks are observed in central Au+Au collisions, consistent with conical flow or Cerenkov radiation.  Discrimination of the two requires further study of the associated $p_T$ dependence.  It should be realized, however, that the combinatorial backgrounds are large in this measurement, and we are investigating potential systematic uncertainties beyond those we have studied.  

% \appendix

% \section{}
% \label{}


\begin{thebibliography}{01}
\bibitem{b1}
J. Adams et al. (STAR Collaboration), Phys. Rev. Lett. {\bf95}, 152301 (2005); 
S.S. Adler et al. (PHENIX Collaboration), Phys. Rev. Lett. {\bf97}, 052301 (2006).

% notes:

% \bibitem{label2}
% If there is a note, it should come last:
% \bibitem{label3} \note
% \end{subbibitems}
\bibitem{b3}
	I. Vitev, Phys. Lett. B {\bf 630}, 78 (2005).
%	\bibitem{b2}
%R. Hwa, presentation at Hard Probes 2006.
\bibitem{b4}
  H. Stoecker, Nucl. Phys. {\bf A750}, 121 (2005); 
  J. Casalderrey-Solana, E. Shuryak and D. Teaney, J. Phys. Conf. Ser. {\bf27}, 23 (2005).
\bibitem{bx}	
	V. Koch, A. Majumder and X.-N. Wang, Phys. Rev. Lett. {\bf 96}, 172302 (2006).
\bibitem{method}
	J. Ulery and F. Wang, nucl-ex/0609016.
\bibitem{b5}
	J. Adams et al. (STAR Collaboration),	Phys. Rev. C {\bf 72}, 014904 (2004).
\bibitem{b6}
	C. Adler et al. (STAR Collaboration), Phys. Rev. C {\bf 66}, 034904 (2002).
\end{thebibliography}
\end{document}